\pdfoutput=1 
\documentclass[a4paper,fleqn]{cas-dc}
\usepackage[authoryear]{natbib}
\usepackage[utf8]{inputenc}
\usepackage{amsmath}
\usepackage{mathtools}
\usepackage{placeins}
\usepackage{subfigure}
\usepackage{epstopdf}
\usepackage{caption}

\begin{document}
\let\WriteBookmarks\relax
\def\floatpagepagefraction{1}
\def\textpagefraction{.001}
\shortauthors{Anis Moradikouchi et~al.}
\pagenumbering{arabic}
\title [mode = title]{ Terahertz Frequency Domain Sensing for Fast Porosity Measurement of Pharmaceutical Tablets}                      

\author[1,2]{Anis Moradikouchi}[
                        orcid=0000-0003-3811-1024]
\cormark[1]

\credit{Conceptualization of this study, Methodology, Software}

\address[1]{Department of Microtechnology and Nanoscience, Chalmers University of Technology, SE-412 96 Gothenburg, Sweden}
\author[2]{Anders Sparén}

\author[3]{Staffan Folestad}
\author[1]{Jan Stake}
\author[1]{Helena Rodilla}

\credit{Data curation, Writing - Original draft preparation}

\address[2]{Oral Product Development, Pharmaceutical Technology \& Development, Operations, AstraZeneca, Gothenburg, Sweden}

\address[3]{Innovation Strategies \& External Liaison, Pharmaceutical Technology \& Development, Operations, AstraZeneca, Gothenburg, Sweden}

\cortext[cor1]{anismo@chalmers.se}

\pagenumbering{arabic}

\shorttitle{}


\begin{keywords}
 \sep Terahertz technology \sep Frequency domain \sep  Non-destructive  \sep Dielectric characterisation\sep Porosity measurements \sep Pharmaceutical tablets
 \sep Tablet sensing
\end{keywords}

\maketitle 
\begin{abstract}
Porosity is an important property of pharmaceutical tablets since it may affect tablet disintegration, dissolution, and bio-availability. It is, therefore, essential to establish non-destructive, fast, and compact techniques to assess porosity, in-situ, during the manufacturing process.
In this paper, the terahertz frequency-domain (THz-FD) technique was explored as a fast, non-destructive, and sensitive technique for porosity measurement of pharmaceutical tablets. We studied a sample set of 69 tablets with different design factors, such as particle size of the active pharmaceutical ingredient (API), Ibuprofen, particle size of the filler, Mannitol, API concentration, and compaction force.
The signal transmitted through each tablet was measured across the frequency range 500-750 GHz using a vector network analyzer combined with a quasi-optical set-up consisting of four off-axis parabolic mirrors to guide and focus the beam. We first extracted the effective refractive index of each tablet from the measured complex transmission coefficients and then translated it to porosity, using an empirical linear relation between effective refractive index and tablet density. The results show that the THz-FD technique was highly sensitive to the variations of the design factors, showing that  filler particle size and compaction force had a significant impact on the effective refractive index of the tablets and, consequently, porosity. Moreover, the fragmentation behaviour of particles was observed by THz porosity measurements and was verified with scanning electron microscopy of the cross-section of tablets.
In conclusion, the THz-FD technique, based on electronic solutions, allows for fast, sensitive, and non-destructive porosity measurement that opens for compact instrument systems capable of in-situ sensing in tablet manufacturing.
\end{abstract}

\section{Introduction}
In the pharmaceutical industry, tablets are one of the most convenient dosage forms for the administration of drugs to patients. Pharmaceutical tablets consist of one or more active pharmaceutical ingredients (API) and excipients. The physical and chemical properties of both API and excipients affect the tableting process and the physical properties of tablets. Porosity, which is the ratio between the air void volume and total bulk volume, is one of them and may play a crucial role in the disintegration and dissolution time of tablets~\citep{Ervasti2012ATablets}. Recently, the pharmaceutical industry has been moving toward continuous manufacturing, and there is a high need for fast and non-destructive techniques to monitor the porosity of tablets during the production process. 
However, there is no in-line technique for monitoring of porosity during
tablet manufacturing~\citep{Naftaly2019IndustrialPlay} and off-line techniques are typically destructive and slow. 

Mercury porosimetry~\citep{Westermarck1998MercuryGranules} and Thermoporometry~\citep{Luukkonen2001InteractionThermoporosimetry} are common reliable methods for porosity measurements with the disadvantage of being destructive.
X-ray computed microtomography~\citep{Markl2017CharacterizationMicrotomography} and helium pycnometry~\citep{Sereno2007DeterminationContent} are reported as non-destructive techniques. X-ray computed microtomography is advantageous for pore structure analysis. Helium pycnometry provides information about average porosity. However, these techniques are slow and not suitable for in-situ sensing, such as for at-line and in-line measurements in manufacturing processing operations. Gas in Scattering Media Absorption Spectroscopy (GASMAS)~\citep{Sjoholm2001AnalysisMedia,Svensson2007NoninvasiveSpectroscopy,Johansson2021OpticalRibbons}
is a non-destructive and at-line technique to measure optical porosity. Dielectric characterisation of tablets using terahertz time-domain spectroscopy (THz-TDS) has recently emerged as a non-destructive tool for porosity measurements~\citep{Lu2020TerahertzReview}. Juuti et al. used a THz spectrometer to demonstrate the correlation between the effective refractive index and tablet porosity~\citep{Juuti2009OpticalTablets}. Bawuah et al. used THz-TDS to obtain the porosity of tablets composed of one excipient (MCC)~\citep{Bawuah2014DetectionTechniques} and later tablets consisting of several excipients and an API~\citep{Bawuah2016TerahertzTablets}. Naftaly et al. proposed the use of THz-TDS coupled with an index-matching material as a destructive technique to analyze
the open porosity and scattering loss of powder compacts~\citep{Naftaly2020MeasuringMedium}.
Skelbæk-Pedersen et al. used  THz-TDS to study particle fragmentation during tableting and showed that initial particle size and deformation behaviour of the particles affect the correlation between effective refractive index and porosity~\citep{Skelbk-Pedersen2020Non-destructiveMeasurements}. Recently, Stanzinger et al. demonstrated a lab-experimental set-up for monitoring powder flow densification, which is inversely proportional to porosity, with THz-TDS~\citep{Stranzinger2019MeasuringSensing}. Whilst THz-TDS instrumentation is a versatile tool for dielectric characterisation in a lab environment, it is difficult to miniaturise and implement as a stable, reliable, and compact porosity sensor system in a production line of pharmaceutical tablets.

In this work, we evaluated the THz frequency-domain (THz-FD) technique, to our knowledge for the first time, as a candidate for a future miniaturized in-line porosity measurement system. The miniaturization capabilities of electronic sources and receivers~\citep{Sengupta2018TerahertzSystems,Hillger2019TerahertzTechnologies}, makes THz-FD promising for a future compact, fast, and non-destructive measurement system~\citep{Neumaier2014MolecularReceiver,Dahlback2012AApproximation}  for in-situ porosity measurements during pharmaceutical processing operations. In our previous work \citep{Moradi2019Non-DestructiveSpectroscopy}, we showed that THz-FD is an effective technique for the characterization of pharmaceutical tablets with different dielectric constants. In this paper, we have explored the applicability of the THz-FD technique to porosity sensing by studying the effect of varying design factors in a tablet formulation on the effective refractive index and, consequently, porosity. The design factors were API and filler particle size, API concentration, and compaction force. The effective refractive index was directly extracted from the measurements of complex transmission coefficients across the frequency range 500-750 GHz. In order to transfer effective refractive index from THz measurements to porosity, an empirical linear relation between effective refractive index and tablet density was used. Moreover, fragmentation behaviour of API and filler particles was observed in THz measurement results and validated with scanning electron microscopy of the tablets.
\section{Methodology}

\subsection{\centering Tablet Preparation}

In this work, we studied a sample set of 69 tablets consisting of Ibuprofen as an API, Mannitol as an excipient, and Magnesium stearate as a lubricant. A full factorial design of API particle size at two levels ($d_{50}$ $\sim$ 71, 154 $\mu$m) and excipient particle size at two levels ($d_{50}$ $\sim$ 91, 450 $\mu$m) was set up, where the $d_{50}$-value is the median diameter of the particle size distribution found at 50\% in the cumulative distribution. This design was repeated at five API concentration levels ($\sim$ 16, 18, 20, 22, and 24 w/w \%), and Magnesium stearate concentration was kept constant at $\sim$ 1 w/w\% for all tablets. Details about the samples are provided in \citep{Sparen2015MatrixSpectroscopy} and a summary is given in the table \ref{tbl1}, where experiments 21–23 are center points. Further information can be found in the table S1 in the supplementary. Tablets from all powder blends were manufactured at three compaction forces ($\sim$ 8, 12, and 16 kN), resulting in 69 different tablet types. All tablets were flat-faced, with a nominal weight of 300 mg, a diameter of 10.0 mm, and a thickness, $l$, in the range of 2.9–3.4 mm measured using a digital micrometer. We used a scanning electron microscope (Zeiss Supra 55), with an accelerating voltage of 5 kV to optically inspect the cross section of one tablet (experiment number 3 in the table \ref{tbl1}). The cross-section of the tablet was coated with a 20-nm film of sputtered aluminium under vacuum.

\begin{table}[width=1.0\linewidth,cols=4,pos=h]
\caption{\textrm {Design of experiments for the tablets. Design factors are API particle size at two levels (71, 154 $\mu$m) , excipient particle size at two levels (91, 450 $\mu$m), API concentration at five levels (16, 18, 20, 22, 25 w/w\% )  and compaction force at three levels (8, 12, 16 kN)  (not included in the table). } }\label{tbl1}

\begin{tabular*}{\tblwidth}{@{} cccc@{} }
\toprule
Experiment & API particle & Filler particle & API concentration\\
number &  size d$_{50}$ ($\mu$m)  & size d$_{50}$ ($\mu$m)& (w/w\%)\\
\midrule

1 & 71 & 91 & 16 \\
2 & 154 & 91 & 16 \\
3 & 71 & 450 & 16 \\
4 & 154 & 450 & 16 \\
5 & 71 & 91 & 18 \\
6 & 154 & 91 & 18 \\
7 & 71 & 450 & 18 \\
8 & 154 & 450 & 18 \\
9 & 71 & 91 & 20 \\
10 & 154 & 91 & 20 \\
11 & 71 & 450 & 20 \\
12 & 154 & 450 & 20 \\
13 & 71 & 91 & 22 \\
14 & 154 & 91 & 22 \\
15 & 71 & 450 & 22 \\
16 & 154 & 450 & 22 \\
17 & 71 & 91 & 24 \\
18 & 154 & 91 & 24 \\
19 & 71 & 450 & 24 \\
20 & 154 & 450 & 24 \\
21 & 95 & 211 & 20 \\
22 & 95 & 211 & 20 \\
23 & 95 & 211 & 20 \\
\bottomrule
\end{tabular*}
  \label{tbl1}
 
\end{table}

\begin{figure}[h!] 
    \centering
        \includegraphics[scale=0.3]{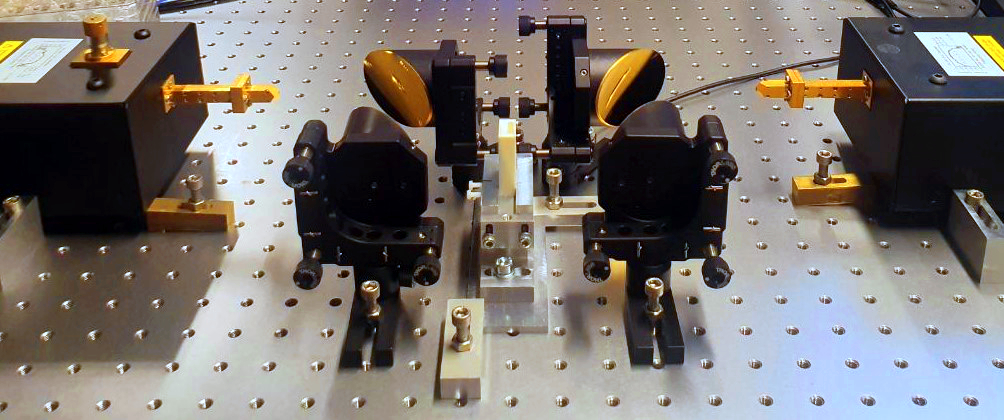}
        \small {(a)}

        \includegraphics[scale=0.70]{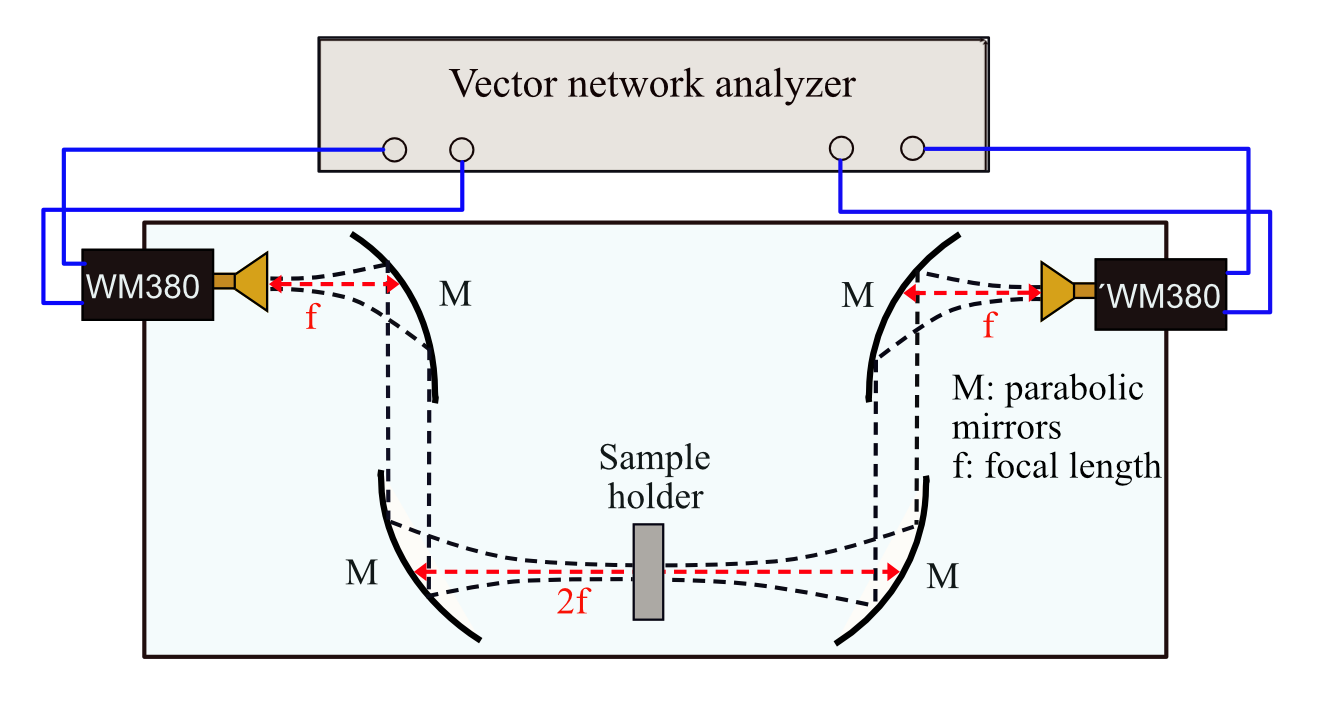}
       \small \\ {(b)}

    \caption{\textrm{a) Photograph  and b) schematics of the experimental set-up showing THz S-parameter measurements in transmission mode. The blue square represents the area showed in the photograph.}}
     \label{fig: 1}
\end{figure}

\subsection{\centering Measurement set-up}

To investigate the tablets, the effective refractive index was extracted from complex transmission coefficients, $S_{21}$, measured at 500-750 GHz and translated into porosity. The measurement set-up consisted of a vector network analyzer (Keysight PNA-X) with VDI frequency extenders WM380 and quasi-optical components, see Fig.~\ref{fig: 1}\href{fig: 1}{a}. The frequency range was selected to avoid scattering effects (long wavelength compared to the particle size), absorption peaks, and at the same time to ensure a small beam size compared to the tablets' diameter. During measurements, the intermediate frequency bandwidth was set to 1 kHz. In order to visualize and help with the alignment of the quasi-optical set-up, the vector network analyzer was calibrated at the interface between the waveguide flanges of the extenders and the horn antennas using the SOLT (Short, Open, Load, Through) method. The signal was then focused on the tablets using four gold protected off-axis parabolic mirrors with a focal length of 76.4 mm, see Fig.~\ref{fig: 1}\href{fig: 1}{b}. The estimated beam diameter at the focal point is approximately 2 mm. Each of the 69 samples was measured ten times in consecutive frequency sweeps to verify the measurement repeatability and estimate the effect of additive white Gaussian noise. Before each tablet measurement, the empty sample holder was measured as a reference for relative measurements. Relative measurements suppressed uncertainty due to possible misalignment of components, loss in the optical path, and temperature and humidity variation. Each frequency sweep took less than one second to complete. The temperature and humidity were recorded during the measurements and varied in the ranges 22.6-22.9 $^{\circ}$C and 21-30 \%, respectively.

\subsection{\centering Effective Refractive Index Extraction} \label{Sec 2.2}
The effective complex refractive  index of  the  samples, $\hat{n}_s=n_s-j\kappa_s$, can be extracted from a relative measurement of the complex transmission coefficient, $T=S_{21}$, between the sample and the empty holder (air), and its comparison to a model~\citep{Zhu2021ComplexRange}.

The relative transmission between sample and air, ${T_s}/{T_a}$, can be modeled as in equation 1, which considers the multiple reflections inside the tablet~\citep{Nicolson1970MeasurementTechniques}.

\begin{equation} \label{eq:1}
  \left(\frac{T_s}{T_a}\right)_m= \frac{(1-\Gamma^2)}{1-\Gamma^2e^{-2jk_0\hat{n}_{s}l}}e^{-jk_0l(\hat{n}_{s}-1)},
\end{equation}
where $k_0$ is the free space wave number, $l$ is the thickness of the tablet and $\Gamma$ is the complex reflection coefficient, which in this case is:

 \begin{equation}\label{eq:2}
   \Gamma=\frac{1-\hat{n}_s}{1+\hat{n}_s}.
\end{equation}

Thick tablets, compared with the wavelength, result in multiple wavelengths inside tablets and therefore phase ambiguity \citep{Bourreau2006AW-band}. In order to address the phase ambiguity and extract $\hat{n}_s$, the steps below were followed. 

Step 1: A linear fit of the phase of measured relative transmission, $\phi = Arg(({T_s}/{T_a})_d)$, across frequency was  used to find the average refractive index as an initial estimate for $n_s$, $n_{s,i}$, using equation \ref{eq:3} \citep{Weir1974AutomaticFrequencies}:

 \begin{equation}\label{eq:3}
   n_{s,i}=1+\frac{\tau c}{2\pi l},
\end{equation} where $c$ is the speed of light in vacuum, and the time delay, $\tau = -\frac{\Delta\phi}{\Delta\omega}$, is the slope of the linear fit~\citep{Mrnka2022AccurateSystems}.

Step 2: The value of $\hat{n}_s$ versus frequency was obtained numerically by solving equation \ref{eq:4}:
 \begin{equation}\label{eq:4}
     f(\hat{n}_s)=\left(\frac{T_s}{T_a}\right)_m-\left(\frac{T_s}{T_a}\right)_d =0.
\end{equation}
This equation has multiple possible solutions because of the periodic nature of the exponential terms. To find the correct solution, the $n_{s, i}$ value from step 1 was used as an initial estimate.

 Step 3: Finally, the model, $({T_s}/{T_a})_m$, was recalculated using $\hat{n}_s$ obtained in step 2, and its magnitude and phase were compared with the measurement data, $({T_s}/{T_a})_d$.
\subsection{\centering Tablet Porosity Calculation}\label{DebsitytoPorosity}

We extracted THz porosity, $f_{THz}$, from the effective refractive index, $n_s$, obtained from the THz measurements.

$f_{THz}$ can be obtained as in~\citep{Sun2004APowders}:
\begin{equation} \label{equ: 5}
 f_{THz}=1-\frac{\rho_{ tab, THz}}{\rho_{true}}, 
\end{equation}
where $\rho_{ tab, THz}$ is the tablet density and $\rho_{true}$ is the true density, which is the density of the particles in the powder formulation. $\rho_{ tab, THz}$ was obtained from the effective refractive index extracted from THz measurements by an empirical linear relation between tablet density and effective refractive index, as in \citep{Moradi2019Non-DestructiveSpectroscopy}:

\begin{equation}\label{equ: 6}
    {\rho_{tab, THz}}= \rho_1 n_s + \rho_0. 
\end{equation}
In order to obtain the coefficients, $\rho_0$ and $\rho_1$, for this specific set of tablets, the tablet density obtained from mechanically measured mass and volume of the samples was used for calibration.

$\rho_{true}$ in equation 5 was calculated from the concentration, $w_i$, and the true density of the ingredients, $\rho_{true, i}$, from literature \citep{Nokhodchi2015CrystalPerformance} and their data sheet, as in \citep{Sun2004APowders}:

\begin{equation}\label{equ: 7}
    \frac{1}{\rho_{true}}=\sum_{i=1}^3\frac{w_i}{\rho_{true, i}}.
\end{equation}
Thus, equations \ref{equ: 5}-\ref{equ: 7} enable us to obtain the porosity of tablets from THz measurements.
\section{Results and Discussion}
\begin{figure}[!htbp]
    \centering
     \hspace*{+0.3cm}
        \includegraphics[scale=0.85]{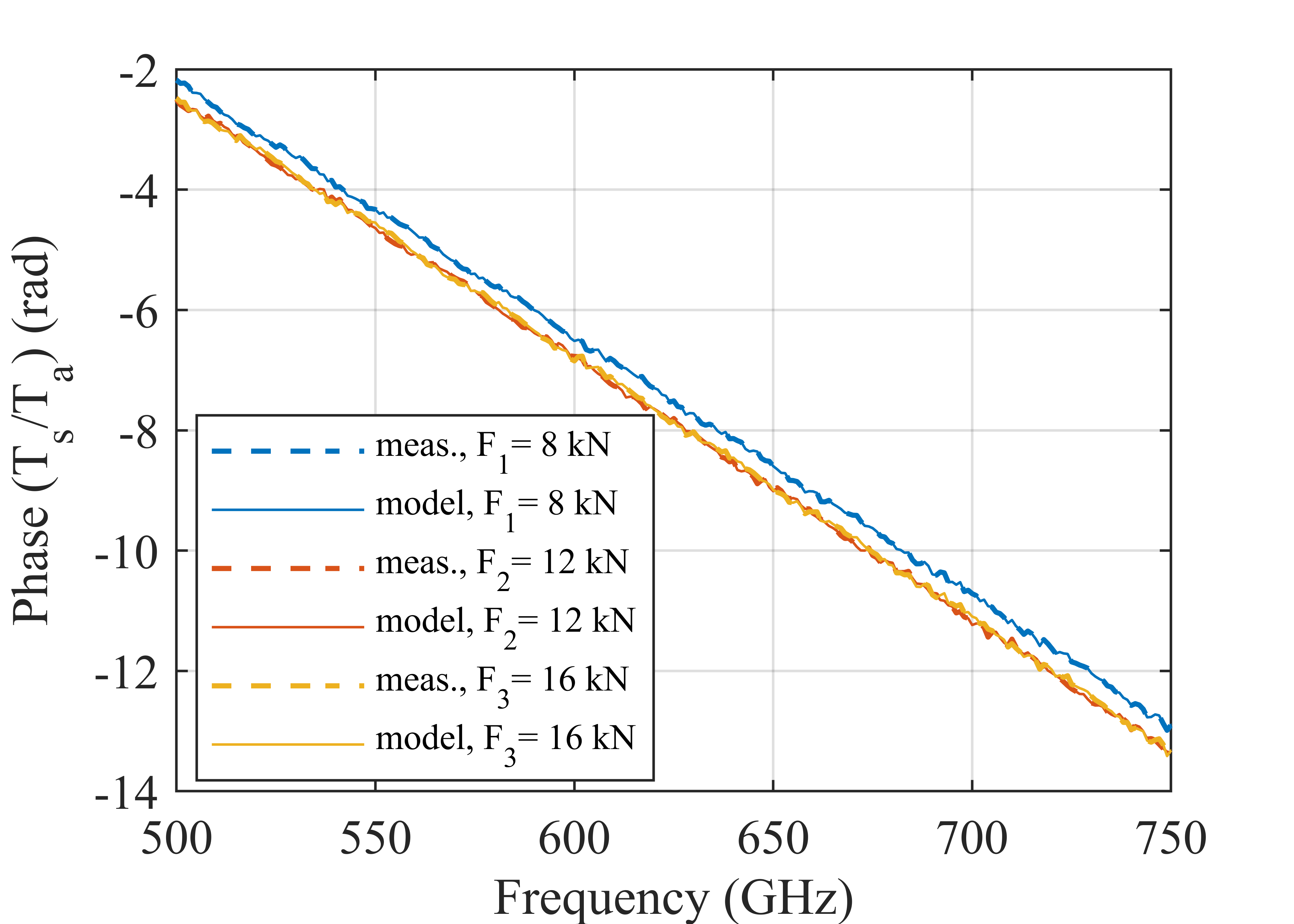}
        \small{\\(a)}
        \includegraphics[scale=0.85]{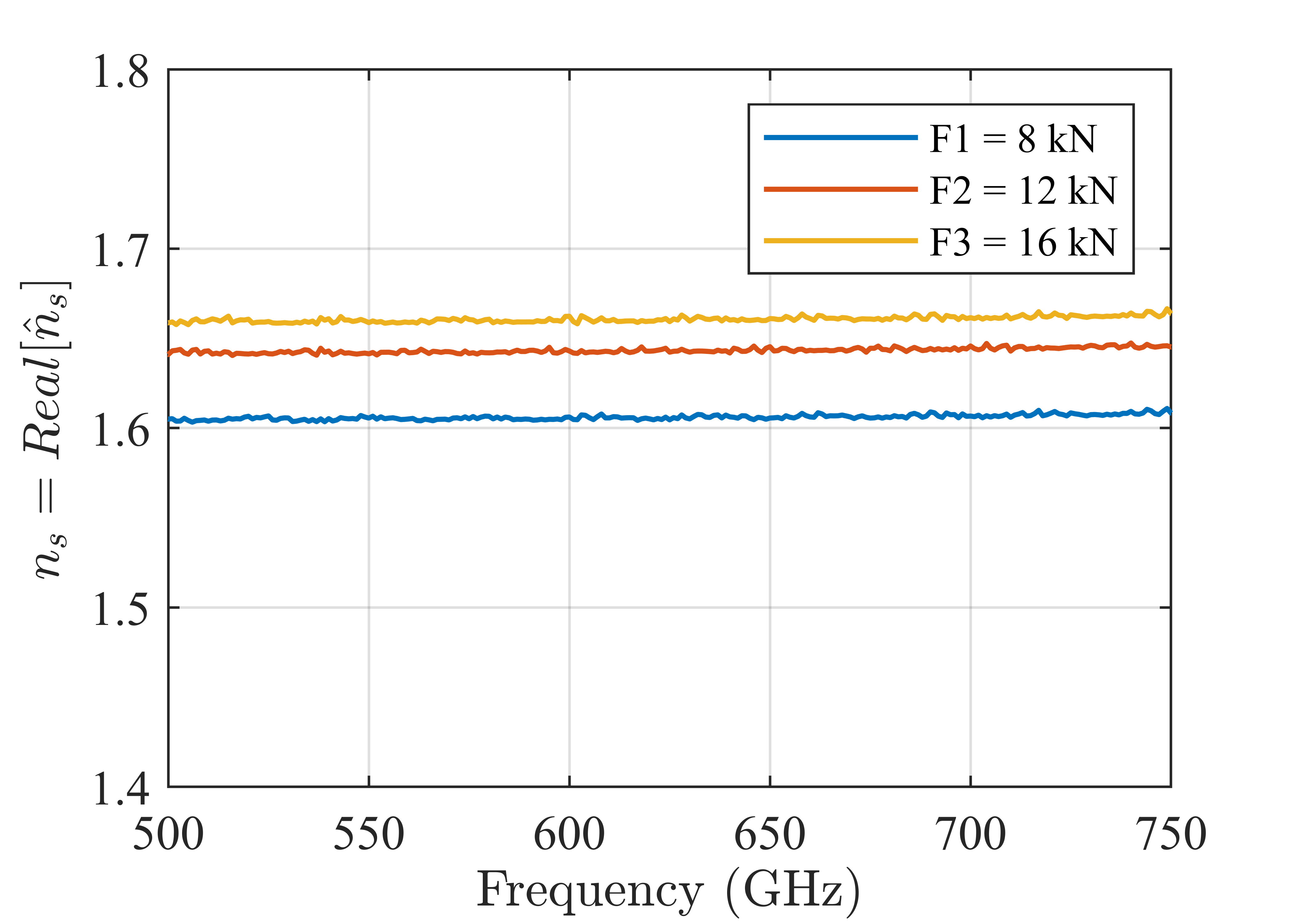}
        
        \small {(b)}
    \caption{\textrm {a) Phase of relative transmission for tablets (experiment number 2 in table \ref{tbl1}) with three compaction forces, $F_1$= 8 kN, $F_2$= 12 kN, $F_3$= 16 kN, showing a good agreement between model and measurement data b) Real part of effective refractive index over frequency. Higher compaction force increased the effective refractive index. }}
    \label{fig: 2}
\end{figure}

We used the real part of the effective refractive index, $n_s$, obtained from the phase of the complex transmission coefficients to characterise the tablets. The computation time required to determine $n_s$ for a complete spectrum consisting of 251 frequency points was about one second. Fig.~\ref{fig: 2}\href{fig: 2}{a} shows the comparison of the measured phase of relative transmission with the model presented in equation \ref{eq:1} for three tablets with the same API and filler particle size, and API concentration (experiment number 2 in table \ref{tbl1}) but three compaction forces, $F_1$= 8 kN, $F_2$= 12 kN, $F_3$= 16 kN. The good agreement with the measurements validates the model. Fig.~\ref{fig: 2}\href{fig: 2}{b} shows the obtained effective refractive index of the three tablets over frequency. 
As can be observed, increasing compaction force increased the effective refractive index because higher compaction force reduces air voids in between tablet particles, increasing the effective refractive index. The minor oscillations observed in $n_s$ were due to standing waves within the measurement set-up and the fact that we used the measured  tablet thickness and not the effective tablet length, which in practice can have small discrepancies due to possible tablet inhomogeneities or beam properties.

 \begin{figure}[!htbp]
  \centering
  \includegraphics[scale=0.85]{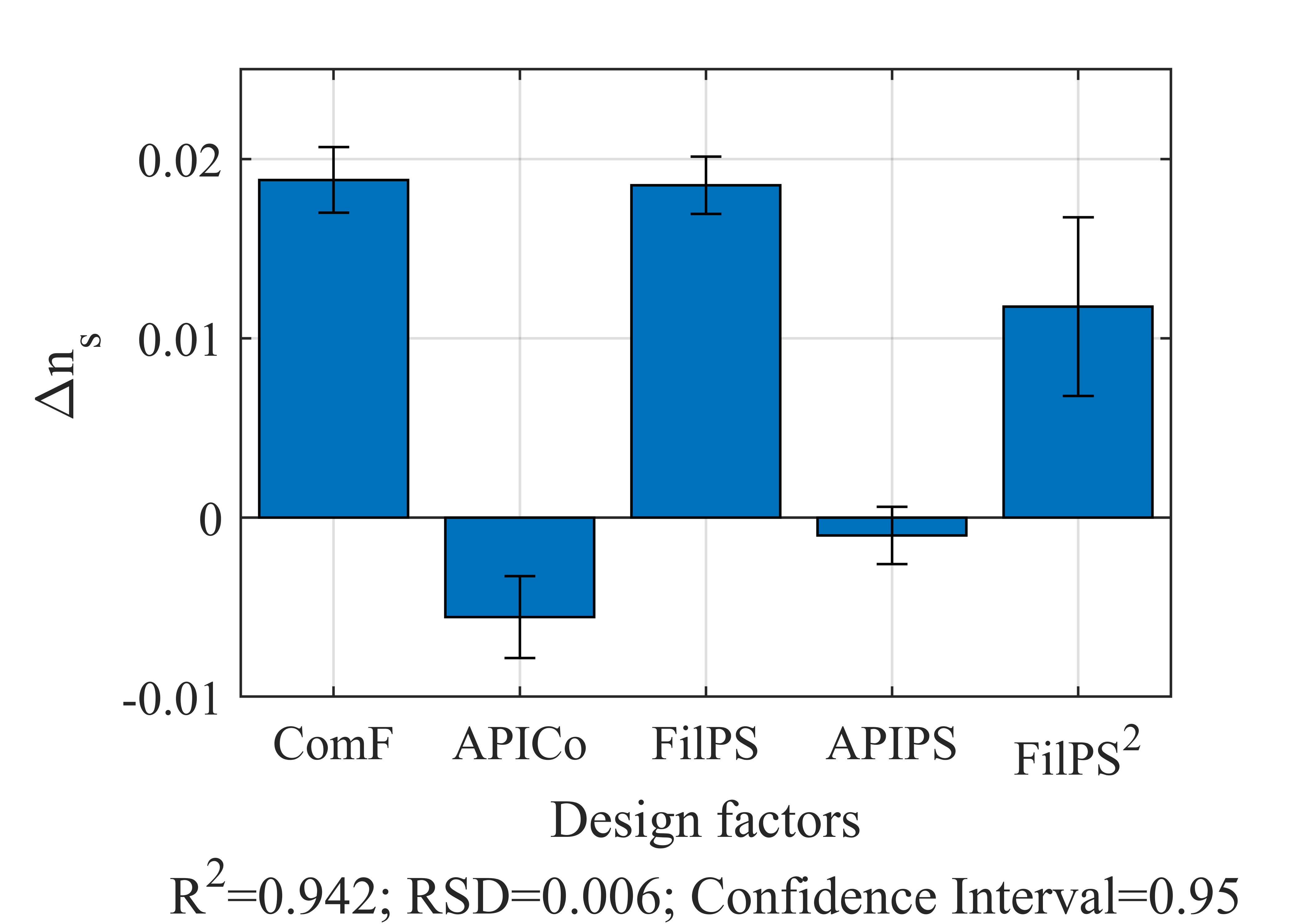}
  \caption{\textrm {Coefficient plot showing the effects of the design factors on the effective refractive index. Compaction force is shown as  ComF, filler particle size as FilPS, API particle size as APIPS, and API concentration as APICo. It can be seen that the filler particle size and compaction force had the largest effects.}}\label{fig: 3}
\end{figure}
\begin{figure} [!htbp]
    \centering
        \includegraphics[scale=0.85]{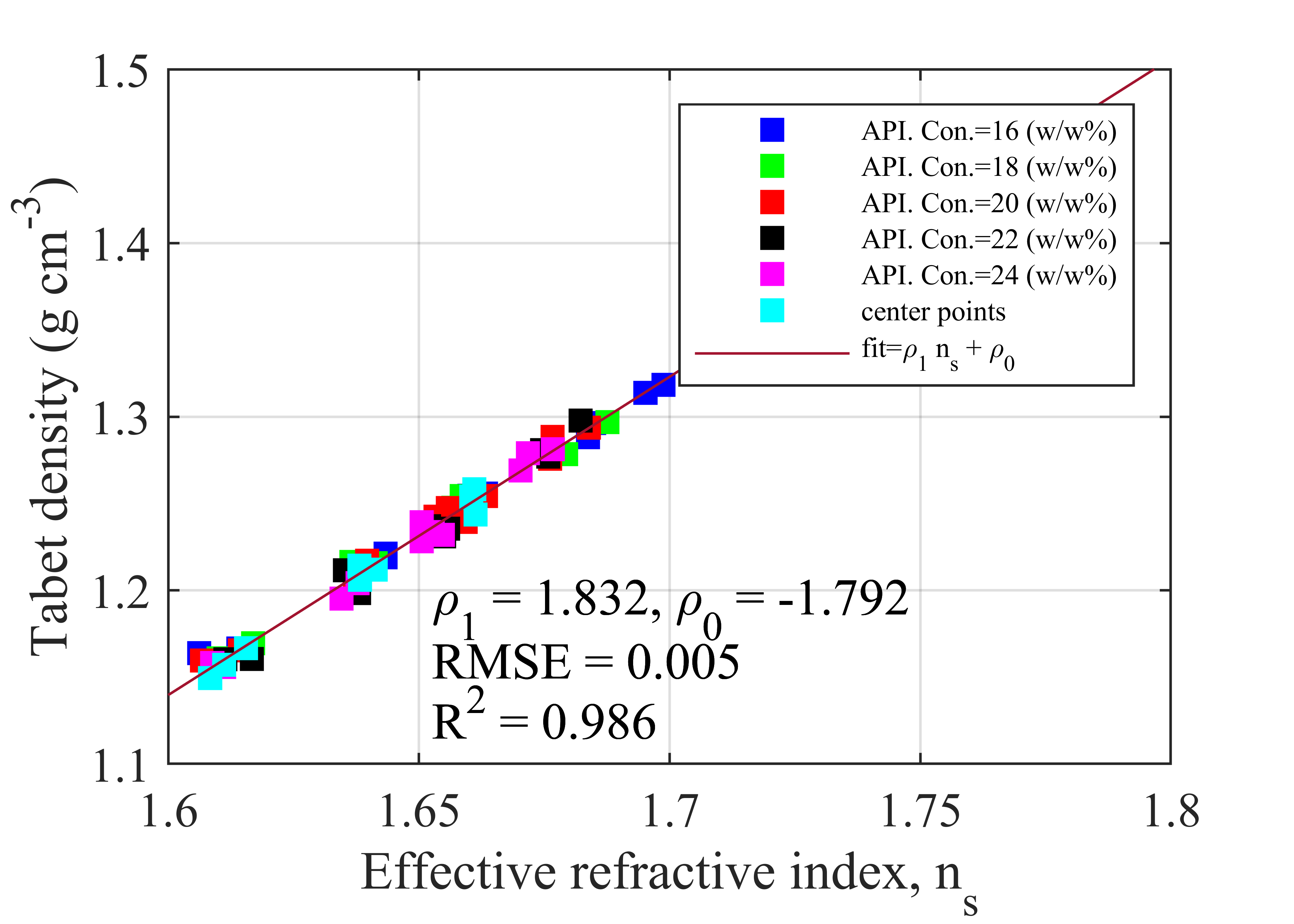}\\
        \small {(a)}  

        \includegraphics[scale=0.85]{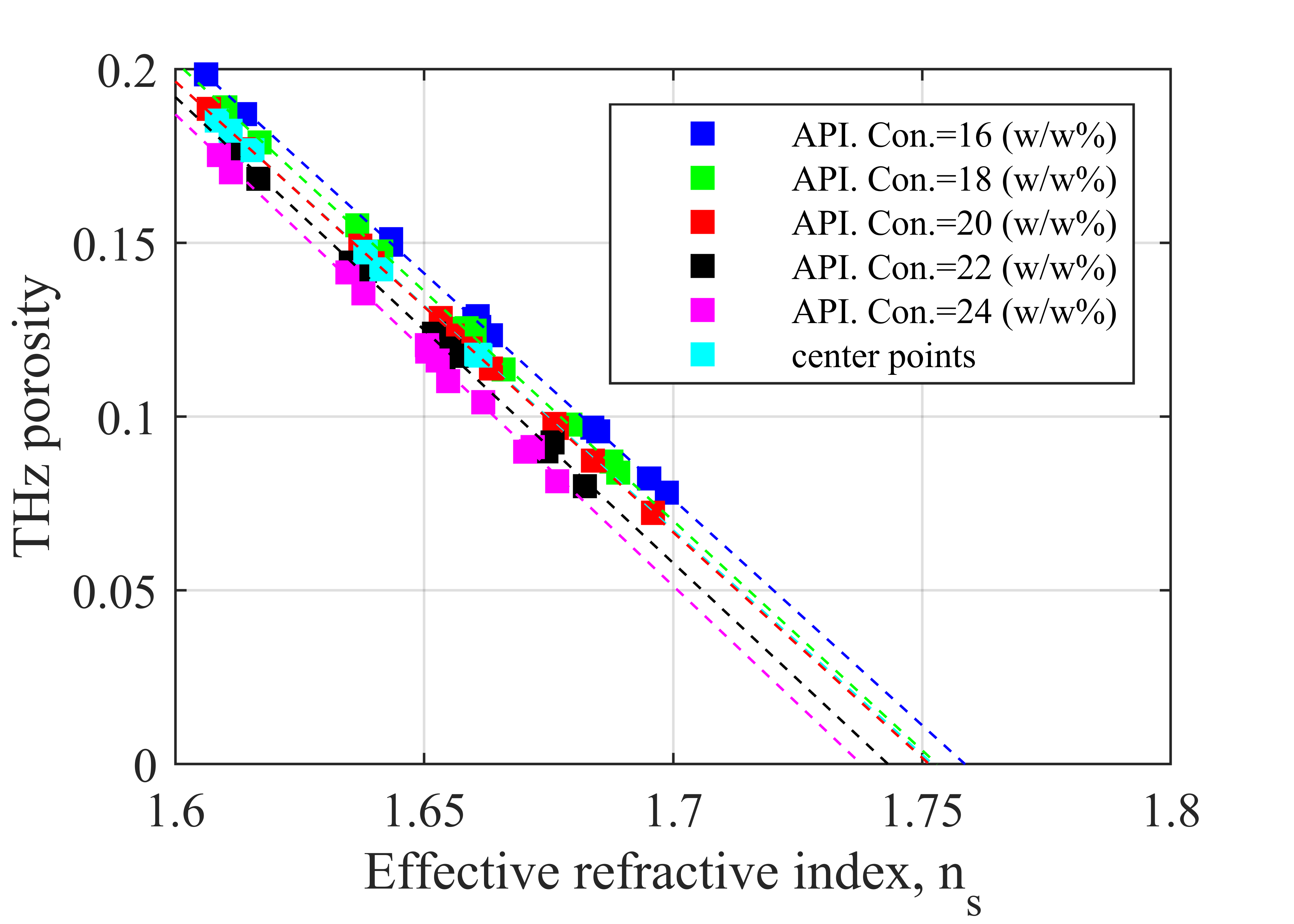}  
       \small \\ {(b)}
 
    \caption{\textrm{a) Correlation between the effective refractive index and tablet density. A linear relationship between the effective refractive index, $n_s$, and tablet density was observed. b) correlation between THz porosity and effective refractive index.}} \label{fig: 4}
\end{figure}
To study the effect of the varied design factors on the effective refractive index of tablets, the mean value of the effective refractive index across the frequency band was chosen as a response. The results were analyzed using a statistical tool for setting up and evaluating the design of experiments (MODDE 12)~\citep{SartoriusStedimDataAnalytics2017User12}, using a multi-linear regression model.

\begin{figure*}
  \centering
  \includegraphics[scale=1.0]{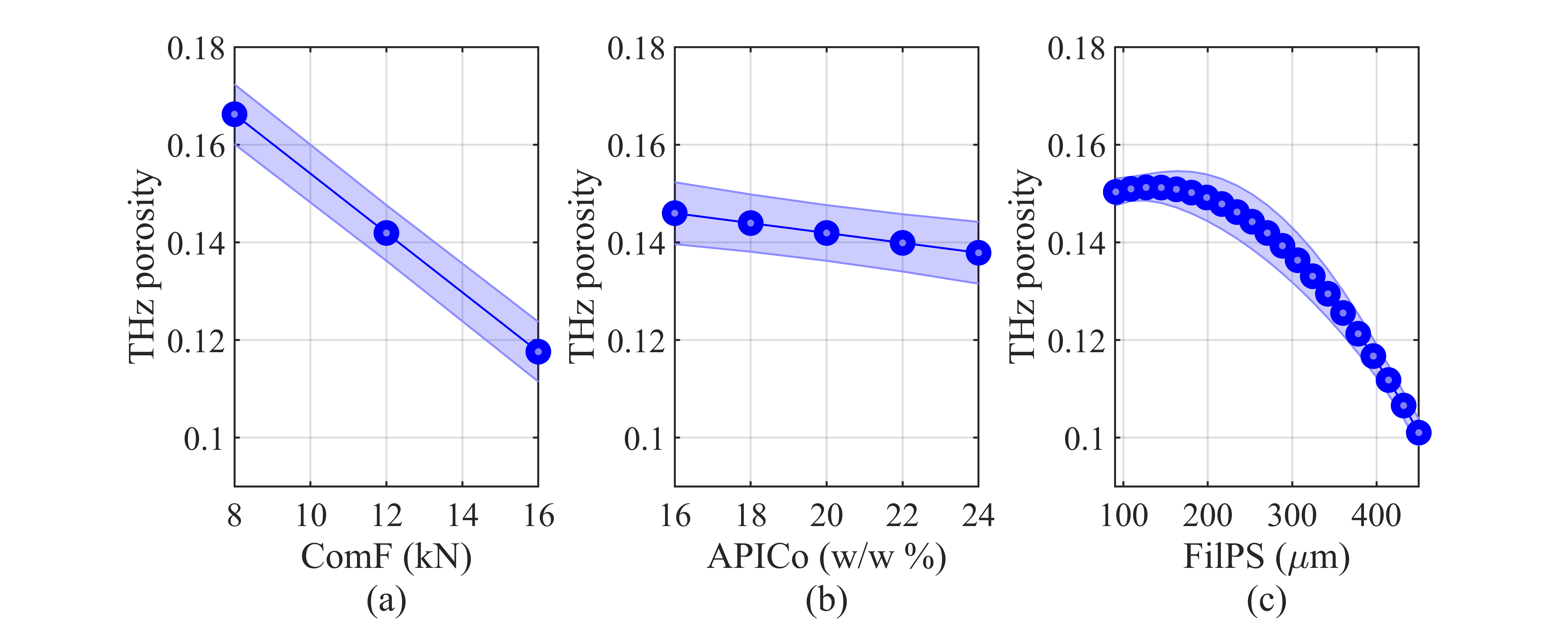}
\caption{\textrm {The variation of THz porosity with respect to design parameters, a) compaction force, b) API concentration, and c) filler particle size are depicted, respectively. Plots are created from fitted first-order and second-order polynomials, similar to equation \ref{equ: 8}. In each plot, the line shows the THz porosity and the shadowed area shows the standard deviation.}}
  \label{fig: 5}
\end{figure*}

Fig.~\ref{fig: 3} shows the coefficient factors (centered and scaled), where the height of the bar represents the change in the effective refractive index, $\Delta n_s$, when a factor varies from its average to high value, having the other factors at their averages. Error bars show the uncertainty in $\Delta n_s$ from each design factor. It can be seen that compaction force, ComF, and filler particle size, FilPS, had the main effects on the effective refractive index. API concentration, APICo, had a minor effect,  and API particle size, APIPS, had a negligible effect on the effective refractive index. Also, the square term of the filler particle size, FilPS$^2$, showed a significant impact, with a positive non-linear effect on the response. The effective refractive index of the tablets can be modeled with multiple linear regression as a function based on the design factors, describing the contribution of each design factor~\citep{DouglasC.Montgomery2000Montgomery:Experiments}. The model is given as in:
\begin{equation} \label{equ: 8}
n_{s}=\beta_{0} + \beta_{1}x_{1} + \beta_{2}x_{2} +\beta_{3}x_{3} + \beta_{33}x_{3}^2 + \hat{e},
\end{equation}
where $x_{1}$ is the compaction force, $x_{2}$ is the API concentration, and $x_{3}$ is the filler particle size. $\beta_0$ is the average value of $n_s$, $\beta_{i}, i=1, 2, 3$ are the corresponding coefficients, and $\hat{e}$ is the error in the model.
 
We used the described linear approximation in section \ref{DebsitytoPorosity} to extract tablet density and porosity from THz measurements. Fig.~\ref{fig: 4}\href{fig: 4}{a} shows the tablet density obtained from mechanically measured mass and volume of the samples versus the obtained effective refractive index. The coefficient of determination, $R^2 = 0.986$, and root mean square error, $\text{RMSE} = 0.005$, show the validity of the linear approximation. 
Moreover, Fig.~\ref{fig: 4}\href{fig: 4}{b} shows the relationship between THz porosity and effective refractive index of the tablets. The refractive index of the solid phase of the tablets for each API concentration, corresponding to zero porosity, can be estimated from the intercept with THz porosity equaling zero.

We extracted THz porosity from the effective refractive index, and studied the effect of the design parameters on the porosity, using the MODDE 12 software. Fig.~\ref{fig: 5} shows the effect of compaction force, API concentration, and filler particle size on the porosity of the tablets. It is seen in Fig.~\ref{fig: 5}\href{fig: 5}{a} that increasing compaction force lowers porosity, which is in agreement with the fact that higher compaction force reduces the air voids in between particles and results in lower porosity. 

Fig.~\ref{fig: 5}\href{fig: 5}{b} shows the effect of API concentration on the porosity. For studying this effect, two factors should be considered. First, according to Fig.~\ref{fig: 3}, increasing Ibuprofen concentration lowered the effective refractive index. Second, Ibuprofen has a lower true density compared with Mannitol, which lowered the total true density. It is seen that as a result, increasing API concentration lowered the porosity of tablets. Moreover, this plot shows the high sensitivity of THz-FD technique to the minor variations of API concentration~\citep{Moradikouchi2021SmallSpectroscopy}. 

\begin{figure}[!htbp] 
  \centering
  \includegraphics[scale=0.28]{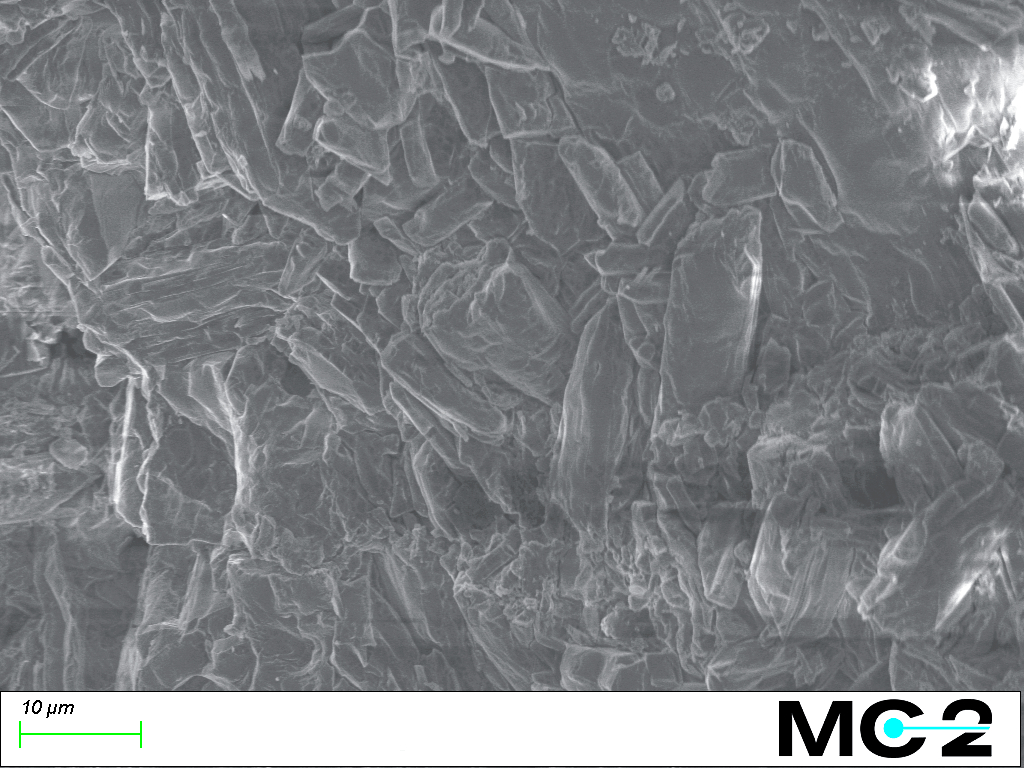}
  \caption{\textrm {Scanning electron microscopy of cross-section of tablet (experiment number 3 in table \ref{tbl1}). The fragmentation of particles while tableting process is clearly seen in this picture.}}
  \label{fig: 6}
\end{figure}

The effect of filler particle size on porosity is shown in Fig.~\ref{fig: 5}\href{fig: 5}{c}. It was observed that increasing particle size of the filler first results in a slight increase in porosity for filler particle size lower than 150~$\mu$m and then a decrease in porosity for larger particles. This behaviour can be explained by particle fragmentation under compaction force during the tableting process. Mannitol \citep{Tarlier2018DeformationSimulator} and Ibuprofen \citep{Liu2008EffectPowders} are needle shape particles, which show friable deformation behaviour under high compression pressure \citep{Tarlier2018DeformationSimulator}, \citep{Le2006InfluenceMethod}. Furthermore, there is a greater tendency for the larger granules to fragment under high compression pressure and therefore create closer packing and consolidation of the particles after fragmentation \citep{Alderborn1985StudiesPressure}, which results in lower porosity. Scanning electron microscopy photographs of the cross-section of one tablet (experiment number 3 in the table \ref{tbl1}) were taken for investigating the fragmentation of the particles. As can be seen in Fig.~\ref{fig: 6}, the results showed a particle size reduction down to around 10~$\mu$m, which verified the fragmentation deformation behavior of Mannitol and Ibuprofen.
\section{Conclusion}
We explored the THz-FD technique for fast porosity measurement of pharmaceutical tablets using a vector network analyzer. The effective refractive index of tablets was extracted from transmission measurements and transformed to the tablet porosity by an empirical linear correlation between tablet density and effective refractive index. We used design of experiments to investigate the applicability of the THz-FD technique to study the effect of factors such as API and filler particle size, API concentration, and compaction force on the porosity. The results showed that the THz-FD technique can detect and quantify changes in the tablet porosity, showing that the filler particle size and compaction force had a major impact on the effective refractive index and, consequently, porosity. Particle fragmentation was observed in the measurement results and verified with scanning electron microscopy. 
The extraction of tablet porosity from VNA measurements can be further refined using all four S-parameters \citep{Bourreau2006AW-band}, and also opens up for calibrated and traceable measurements \citep{Ridler2019StrategiesAnalyzer}. This work showed that the THz-FD technique is promising for fast, non-destructive, and sensitive porosity measurements of pharmaceutical tablets. Moreover, the possibility to miniaturise terahertz sensors based on electronic solutions opens up for future compact porosity measurement systems.

\section*{Declaration of Competing Interest}
The authors declare that they have no known competing financial interests or personal relationships that could have appeared to influence the work reported in this paper.

\section*{Acknowledgment}
The authors would like to thank Mr. Mats Myremark for
machining parts for the measurement setup. The measurements were carried out at the Kollberg Laboratory, at Chalmers
University of Technology, Gothenburg, Sweden.
\section*{Funding}
This research was funded by a grant from the Swedish foundation for strategic research (SSF), ID 16-0050, and from AstraZeneca, Gothenburg, Sweden.

\newcommand*{\doi}[1]{\href{https://doi.org/\detokenize{#1}}{ \detokenize{#1}}}
\bibliographystyle{cas-model2-names}
\bibliography{references}

\end{document}


\begin{table*}[t]
\label{tbl1}

\caption{ \textbf{Table S1:} List of the characterized tablets in this study. The ingredients were Ibuprofen supplied by IOL Chemicals and Pharmaceuticals (Punjab, India), Mannitol Parteck M100 and Mannitol Parteck M200 from Merck KGaA (Darmstadt, Germany), Mannitol Pearlitol 400DC from Roquette (Lestrem, France), Magnesium stearate from AstraZeneca R\&D (Södertälje, Sweden) . The thickness of each tablet was measured five times by a micrometer. Tablet density was calculated from the weight and dimensions of tablets, and the true density of the compact powder was calculated using equation 7.}

\begin{tabular*}{\textwidth}{@{\extracolsep{\fill}} ccccccc}
\toprule
\toprule
Tablet &  Thickness & Mass & Tablet density  & True density & API concentration & Compaction force \\
 & (mm) & (mg) & (g cm$^-3$)  & (g cm$^-3$) & (w/w \%) & (kN) \\
\midrule
\textbf{Set 1} \\
1 & 3.340 & 306 & 1.166 & 1.433 & 16.8  & 7.4 \\
  & 3.190 & 306 & 1.221 &  & 16.8  & 12.2\\
   & 3.106 & 306 & 1.254 &  & 16.8  & 16.0 \\
  
2 & 3.325 & 304 & 1.164 & 1.436 & 16.2  & 7.9 \\
 & 3.175 & 304 & 1.219 &  & 16.2  & 12.1 \\
  & 3.089 & 304 & 1.253 &  & 16.2  & 15.9 \\
   
3 & 3.072 & 303 & 1.256 & 1.432& 16.0 & 8.0 \\
 & 2.976 & 303 & 1.296 & & 16.0 & 11.8 \\
 & 2.926 & 303 & 1.318 &  & 16.0 & 15.8 \\
 
4 & 3.095 & 305 & 1.255 & 1.431& 16.2  & 8.4 \\
& 3.015 & 305 & 1.288 & & 16.2  & 11.9 \\
& 2.956 & 305 & 1.314 &  & 16.2  & 15.9 \\
\midrule
\textbf{Set 2} \\

5 & 3.298 & 303 & 1.170 & 1.425 & 18.5  & 8.2 \\
& 3.174 & 303 & 1.216 &  & 18.5  & 12.2 \\
& 3.076 & 303 & 1.254 &  & 18.5  & 15.8\\

6 & 3.356 & 306 & 1.161 & 1.427 & 18.0  & 7.9 \\
& 3.203 & 306 & 1.216 &  & 18.0  & 12.0 \\
& 3.121 & 306 & 1.248 &  & 18.0  & 15.9\\

7 & 3.123 & 306 & 1.248 & 1.421 & 18.4  & 8.2 \\
& 3.029 & 306 & 1.286 &  & 18.4  & 12.1 \\
& 2.961 & 306 & 1.316 &  & 18.4  & 15.8\\

8 & 3.117 & 307 & 1.254 & 1.424 & 17.9  & 8.5 \\
& 3.057 & 307 & 1.278 &  & 17.9 & 11.9 \\
& 3.014 & 307 & 1.297 &  & 17.9  & 15.3\\ 

\midrule
\textbf{Set 3} \\
9 & 3.321 & 304 & 1.165 & 1.416 & 20.6  & 8.2 \\
& 3.180 & 304 & 1.217 &  & 20.6 & 12.2 \\
& 3.103 & 304 & 1.247 &  & 20.6  & 16.0\\ 

10 & 3.371 & 307 & 1.160 & 1.419 & 19.9  & 7.8 \\
& 3.250 & 307 & 1.203 &  & 19.9 & 11.7\\
& 3.146 & 307 & 1.242 &  & 19.9  & 15.4\\ 

11 & 3.066 & 302 & 1.254 & 1.417 & 19.5  & 8.0 \\
& 2.985 & 302 & 1.288 &  & 19.5 & 11.9 \\
& 2.972 & 302 & 1.294 &  & 19.5  & 15.2\\ 

12 & 3.144 & 306 & 1.239 & 1.418 & 19.3  & 8.4 \\
& 3.054 & 306 & 1.276 & & 19.3 & 12.0\\
& 3.008 & 306 & 1.295 &  & 19.3  & 16.0\\

\midrule
\textbf{Set 4} \\
13 & 3.335 & 304 & 1.161 & 1.407& 22.7  & 8.3 \\
& 3.195 & 304 & 1.212 &  & 22.7& 11.9 \\
& 3.133 & 304 & 1.236 &  & 22.7  & 16.1\\ 

14 & 3.336 & 304 & 1.160 & 1.409& 22.2  & 8.3 \\
& 3.230 & 304 & 1.198 &  & 22.2 & 11.8 \\
& 3.129 & 304 & 1.237 &  & 22.2  & 15.6\\ 

15 & 3.124 & 303 & 1.235 & 1.402 & 22.9 & 7.9 \\
& 3.012 & 303 & 1.281 &  & 22.9 & 13.0 \\
& 2.973 & 303 & 1.298 &  & 22.9  & 16.0\\ 

\end{tabular*}
\end{table*}

\begin{table*}[t]
\begin{tabular*}{\textwidth}{@{\extracolsep{\fill}} ccccccc}
\toprule
Tablet &  Thickness & Mass & Tablet density  & True density & API concentration & Compaction force \\
 & (mm) & (mg) & (g cm$^-3$)  & (g cm$^-3$) & (w/w \%) & (kN) \\
\midrule
16 & 3.154 & 305 & 1.231 & 1.408& 21.4 & 7.9 \\
& 3.071 & 305 & 1.264 &  & 21.4  & 11.7\\
& 3.041 & 305 & 1.277 &  & 21.4   & 14.6\\ 
\midrule
\textbf{Set 5} \\
17 & 3.348 & 304 & 1.156 & 1.398 & 24.8  & 7.8 \\
  & 3.215 & 304 & 1.204 &  & 24.8  & 11.8\\
   & 3.151 & 304 & 1.228 &  & 24.8  & 15.5 \\
  
18 & 3.375 & 307 & 1.158 & 1.401 & 24.1  & 8.0 \\
 & 3.270 & 307 & 1.195 &  & 24.1   & 11.5 \\
  & 3.154 & 307 & 1.239 &  & 24.1   & 15.8\\
   
19 & 3.142 & 304 & 1.232 & 1.393 & 25.0 & 7.8 \\
 & 3.050 & 304 & 1.296 & & 25.0 & 12.1 \\
 & 3.021 & 304 & 1.281 &  & 25.0 & 15.6 \\
 
20 & 3.148 & 305 & 1.234 & 1.398 & 23.8  & 8.5 \\
& 3.064 & 305 & 1.267 &  & 23.8  & 12.3 \\
& 3.036 & 305 & 1.279 &  & 23.8  & 16.0 \\
\bottomrule
\textbf{Center points} \\
21 & 3.372 & 309 & 1.167 & 1.419 & 20.0  & 8.7 \\
& 3.240 & 309 & 1.214 &  & 20.0 & 12.3 \\
& 3.164 & 309 & 1.244 &  & 20.0  & 16.1 \\

22 & 3.378 & 307 & 1.157 & 1.418 & 20.1  & 8.3 \\
& 3.241 & 307 & 1.206 &  & 20.1 & 11.8 \\
& 3.107 & 307 & 1.258 &  & 20.1  & 16.7 \\

23 & 3.378 & 305 & 1.149 & 1.417 & 20.4  & 7.9 \\
& 3.204 & 305 & 1.212 &  & 20.4 & 12.0 \\
& 3.097 & 305 & 1.254 &  & 20.4  & 15.9 \\

\bottomrule
\bottomrule
\end{tabular*}
\end{table*}